
\input harvmac
\font\ninerm=cmr9
\def\figfont{\ninerm}
\def\pictures{y }
\message{ Include the PiCTeX figures (y/n)? }\read-1 to\pansw
\ifx\pansw\pictures\message{(Figures will be included).}
\input pictex
\else\message{(Figures will not be included).}
\fi

\def\spur#1{\mathord{\not\mathrel{#1}}}

\def\lte{\mathrel{\displaystyle\mathop{\kern 0pt <}_{\raise .3ex
\hbox{$\sim$}}}}
\def\gte{\mathrel{\displaystyle\mathop{\kern 0pt >}_{\raise .3ex
\hbox{$\sim$}}}}

\def\cA{{\cal A}}

\def\cD{{\cal D}}

\def\cG{{\cal G}}

\def\cL{{\cal L}}

\def\cO{{\cal O}}

\def\e{\hbox{e}}

\newdimen\tdim
\tdim=1pt
\def\stpltsmbl{\setplotsymbol ({\sevenrm .})}
\def\tarrow{\arrow <5\tdim> [.3,.6]}
%
\def\hfloop{\beginpicture
    \setcoordinatesystem units <\tdim,\tdim>
    \setplotsymbol ({\tenrm .})
    \circulararc 360 degrees from 0 14 center at 0 0
    \circulararc 360 degrees from 0 15 center at 0 0
\endpicture}
%
%
\def\lfloop{\beginpicture
    \setcoordinatesystem units <\tdim,\tdim>
    \circulararc 360 degrees from 0 15 center at 0 0
\endpicture}
%
%
\def\lebox{ \beginpicture
    \setcoordinatesystem units <\tdim,\tdim>
    \multiput {.} at -16  16 *8  4  0 /
    \multiput {.} at  16  16 *8  0 -4 /
    \multiput {.} at  16 -16 *8 -4  0 /
    \multiput {.} at -16 -16 *8  0  4 /
\endpicture}

\def\phru{ \beginpicture
\setcoordinatesystem units <\tdim,\tdim> point at 2 0
\stpltsmbl
\setquadratic
\plot
0 0
2.5 3
5 0
7.5 -3
10 0
/
\endpicture}

\def\phrd{
\beginpicture
\setcoordinatesystem units <\tdim,\tdim> point at 2 0
\stpltsmbl
\setquadratic
\plot
0 0
2.5 -3
5 0
7.5 3
10 0
/
\endpicture}

\def\phdr{
\beginpicture
\setcoordinatesystem units <\tdim,\tdim> point at 2 0
\stpltsmbl
\setquadratic
\plot
0 0
3 -2.5
0 -5
-3 -7.5
0 -10
/
\endpicture}

\def\jagru{ \beginpicture
\setcoordinatesystem units <\tdim,\tdim> point at 2 0
\stpltsmbl
\plot
0 0
2.5 3
5 0
7.5 -3
10 0
/
\endpicture}

\def\jagrd{
\beginpicture
\setcoordinatesystem units <\tdim,\tdim> point at 2 0
\stpltsmbl
\plot
0 0
2.5 -3
5 0
7.5 3
10 0
/
\endpicture}

\def\jagdr{
\beginpicture
\setcoordinatesystem units <\tdim,\tdim> point at 2 0
\stpltsmbl
\plot
0 0
3 -2.5
0 -5
-3 -7.5
0 -10
/
\endpicture}

\def\sru{
\beginpicture
\setcoordinatesystem units <\tdim,\tdim> point at 2 0
\stpltsmbl
\setquadratic
\plot
  0.0   0.0
  4.8   1.5
  7.5   5.0
  7.3   8.5
  5.0  10.0
  2.7   8.5
  2.5   5.0
  5.2   1.5
 10.0   0.0
/
\endpicture}

\def\srd{
\beginpicture
\setcoordinatesystem units <\tdim,\tdim> point at 2 0
\stpltsmbl
\setquadratic
\plot
  0.0   0.0
  4.8  -1.5
  7.5  -5.0
  7.3  -8.5
  5.0 -10.0
  2.7  -8.5
  2.5  -5.0
  5.2  -1.5
 10.0  -0.0
/
\endpicture}

\def\sdr{
\beginpicture
\setcoordinatesystem units <\tdim,\tdim> point at 2 0
\stpltsmbl
\setquadratic
\plot
  0.0   0.0
  1.5  -4.8
  5.0  -7.5
  8.5  -7.3
 10.0  -5.0
  8.5  -2.7
  5.0  -2.5
  1.5  -5.2
  0.0 -10.0
/
\endpicture}

\Title{HUTP-93/A021}{Effective Field Theories with Instantons\footnote{$^*$}
{Research supported in
part by the National Science Foundation,
under grant \#PHY-9218167, and in part by the Texas National
Research Laboratory Commission under grant \#RGFY93-278B.}}

\centerline{Howard Georgi\footnote{$^*$}
{Research supported in
part by the National Science Foundation,
under grant \#PHY-9218167, and in part by the Texas National
Research Laboratory Commission under grant
\#RGFY93-278B.}\footnote{$^\dagger$}
{(georgi@huhepl.harvard.edu)}
 and Samuel T. Osofsky\footnote{$^{*\dagger\dagger}$}
{(osofsky@huhepl.harvard.edu)}}
\bigskip\centerline{Lyman Laboratory of Physics}
\centerline{Harvard University}\centerline{Cambridge, MA 02138}


\vskip .3in
We explain how nonperturbative effects can be systematically included
when constructing an effective field theory.  We concentrate in particular
on the matching calculation, by which a low energy theory without a
heavy particle degree of freedom is matched onto the full theory. The
matching calculation requires some new formalism, because for a large
class of theories, instantons occur in a continuous range of scales from
far below the mass of the heavy particle to far above.  We show how the
complete IR finiteness of the usual matching formalism arises in this new
context.  As an example, we formally integrate out a heavy quark in QCD.

\Date{07/93} 


\nfig\fLEBOX{}
\nfig\fFEYNMAN{}
\nfig\fPROD{}
\nfig\fSUMEM{}
\nfig\fSHRINKEM{}
\nfig\fANTI{}
\nfig\fQCDFEYN{}
\nfig\fQCDINTS{}
\nfig\fDLONEZERO{}
\nfig\fDLZEROONE{}
\nfig\fEXPLONE{}
\nfig\fEXPLTWO{}
\nfig\fEXPLTHREE{}
\nfig\fDLONEONE{}

\newsec{Introduction}

In constructing an effective field
theory\foot{
    For a review of effective field theory, see
    \ref\rHGREV{H. Georgi, HUTP-93/A003, to appear in Ann. Rev. Nucl.
        Part. Sci., {\bf 43}.
    }.
}
from a more fundamental theory,
one starts with the renormalization scale $\mu$ at some high scale and
then uses the renormalization group to evolve $\mu$ continuously down
to lower and lower scales, eventually encountering the mass scale $M$ of
one of the particles in the theory.  At scales below $M$, perturbative
arguments imply decoupling: in a physical renormalization scheme,
graphs with virtual heavy particles are suppressed by powers of $p/M$,
where $p$ is an external momentum in the low energy
theory \ref\rAPCAR{T. Appelquist and J. Carazone, Phys. Rev. {\bf D11} (1975)
2856.}.

However, for tractable evaluation of higher loop
diagrams, and in order to simplify the renormalization group
equations, a mass independent scheme such as MS is more attractive
than a physical scheme.
In a mass independent scheme, decoupling must
be included by hand: after $\mu$ has been evolved to $M$, the theory is
matched onto a low energy theory where the only degrees of freedom are
the light fields.  The matching is achieved order by order in some small
parameter, such as in a loop expansion.  The new theory is then evolved
via the renormalization group to lower scales.

In this paper we investigate how instanton effects can be included in
this picture, in particular in the matching calculation, since we find
that the continuous part of the renormalization group is only
trivially modified. There are two very different reasons for doing this.

In a large class of theories, instantons occur in a
continuous range of scales from far below $M$ to far above.  By this
we mean that the functional integral is dominated by solutions (or
approximate solutions) of the
bosonic Euclidean equations of motion, and the bosonic action is
classically scale invariant over that range.
However,
the effective field theory is necessarily an expansion to
some finite order $n$ in $p/m$, where $p$ is a momentum in the low
energy theory.  For fixed $n$ the effective theory will duplicate the
physics of the full theory (to some accuracy) when $p/m < 1$, but
its behavior will be wildly different when $p > m$.  This leads to
the first motivation for studying nonperturbative contributions to
matching: to determine the role of instantons, and in particular the
role of instantons smaller than $1/m$, in the low energy theory.
These small instantons necessarily involve momentum scales $p$ with
$p > m$.  We will show that, as might be expected, the small
instantons can very naturally be included in the matching corrections,
and the large ones left in the effective theory.  To do the matching
calculation we will find it useful to think of the dilute gas
approximation as a map from Lagrangians to Lagrangians, and to
simultaneously expand in instanton density and the number of loops.

The second reason is that matching should provide one
of the few calculations involving instantons (in a classically scale
invariant, asymptotically free theory) which does not suffer from IR
problems. These problems arise for the same reason
perturbative calculations at small momenta fail in such theories,
although in this case it is large instantons that create the problem:
for low energy scales the coupling grows too large for perturbation
theory, or a semiclassical expansion, to be trusted.  But a matching
calculation should have no IR problems, because all the IR physics
should cancel.  In particular, the dilute gas approximation, by
which we will mean a perturbative expansion in instanton density at
the heavy fermion scale, will be trustworthy, if instantons at that
scale and above are dilute.  This is in contrast to most instanton
calculations, where large instantons ruin the calculation whatever
scale it is performed at.  We will see this IR cancellation in a
specific example.

The remainder of the paper is organized as follows.  In sect. 2, we
review effective field theories.  In sect. 3 we briefly review
instantons and comment on renormalization and the renormalization
group in that context.  In sect. 4 we explain the matching
calculation.  In sect. 5 we formally apply the method of sect. 4 to
matching at a heavy quark threshold in QCD.  Finally, in sect. 6 we
comment about practical calculations.

\newsec{Effective Field Theories}

In this section, we briefly review the formalism of effective field theories,
following \rHGREV.

Given a quantum field theory (whether effective or not), a low energy
effective field theory can
be constructed as follows.  Begin at a very large scale, that is, with the
renormalization scale, $\mu$, very large.  In a strongly interacting
theory or a theory with unknown physics at high energy, this starting
scale should be sufficiently large that nonrenormalizable interactions
produced at higher scales are too small to be relevant.  In a
renormalizable, weakly interacting theory, $\mu$ should be above the
masses of all the particles, where the effective theory is given
simply by the renormalizable theory, with no nonrenormalizable terms.

In this region, the physics is described by a set of fields, $\Phi$,
describing the heaviest particles, of mass $M$, and a set of light
particle fields, $\phi$, describing all the lighter particles.  The
Lagrangian has the form
\eqn\eHL{
    \cL_H(\Phi,\phi) + \cL_L(\phi)
}
where $\cL_L(\phi)$ contains all the terms that depend only on the
light fields, and $\cL_H(\Phi,\phi)$ is everything else.  The theory
is then evolved down to lower scales.  As long as no particle masses
are encountered, this evolution is described by the renormalization
group.  However, when $\mu$ drops below the mass, $M$, of the heavy
particles, decoupling should be put in by hand by switching to an effective
theory
without the heavy particles. In the process, the parameters of the
theory change, and new interactions, some nonrenormalizable, may be
introduced. The Lagrangian of the effective theory below $M$ has
the form
\eqn\eLELAG{
  \cL_L(\phi) + \delta \cL(\phi),
}
where $\delta \cL(\phi)$  is the ``matching correction" that contains
the new interactions.

Both the changes in existing parameters, and the coefficients of the
new interactions are computed by ``matching" the physics just below the
boundary in the two theories. Matching determines the sizes of the
nonrenormalizable terms associated with the heavy particles.
Generally, the only way we have of calculating matching corrections is
in perturbation theory in some small parameter.  It can be done in
a loop expansion which, if there are no factors of $\hbar$ hidden
in the parameters of the Lagrangian, is equivalent to an expansion in
powers of $\hbar$,
\eqn\eLOOPEXP{
  \delta \cL = \sum^{\infty}_{l=0} \; \delta \cL^{(l)}.
}
The matching correction, $\delta \cL(\phi)$, should be chosen so that
the physics of the $\phi$ particles is the same in the two theories at
$\mu = M$, just at the boundary.

The strongest form of this equivalence that we might impose is that
all one-light-particle irreducible (1LPI) functions with external light
particles are the same.  By 1LPI is meant a graph that cannot be
disconnected by cutting a single {\it light} particle line.  We can
describe this in terms of the equality of the light particle effective
actions in the two descriptions.  The light particle effective action
in the low energy theory is just the effective action.  In the high
energy theory, the light particle effective action is defined as
follows:  construct the generating functional for connected Feynman
graphs with light particles external lines, and then Legendre
transform it.  This gives the generating functional for 1LPI graphs.
Then the matching condition is\foot{A weaker condition could be
imposed -- that the $S$-matrix for light particle scattering is the
same in the two theories. There are arguments (which
we do not understand) that in some
cases only this weaker form of equivalence can be imposed: attempting
to match the greens functions, rather then just the $S$-matrix, leads
to a nonlocal effective field
theory \ref\rSTEVE{Stephen D.H. Hsu, HUTP-93.}.}
\eqn\eHGMATCH{
  S[\cL_H + \cL_L] = S[\cL_L + \delta \cL].
}
We have changed the notation in \eHGMATCH\ from that of \rHGREV\ in two ways
for typographical convenience. We have suppressed the dependence on the
classical field, $\phi$, and we have indicated the dependence on the
Lagrangian by square brackets.

Formally, we imagine expanding $S[\cL]$, where
$\cL$ is an arbitrary Lagrangian, in the number of loops $l$:
\eqn\eSLOOPEXP{
    S[\cL] = \sum_l S^{(l)}[\cL]
}
It should be stressed that $S^{(l)}$ is a projection of $S$ onto the
$l$-loop part.  If there are terms from loop effects already present
in $\cL$, then the loop-order of these terms must be taken into
account when calculating $S^{(l)}$.

At tree level, the light particle effective action in the low energy
theory is just the Lagrangian itself, thus
\eqn\HGTREE{
  S^{(0)}[\cL_L + \delta \cL^{(0)}] = \int \cL_L  + \int \delta \cL^{(0)}.
}
In the high energy theory, the light particle effective action is the
sum of all 1LPI tree graphs.  Thus it is a sum of the light particle
Lagrangian plus all trees with external light particles and only
internal heavy particle lines:
\eqn\HGTREEI{
  S^{(0)}[\cL_H + \cL_L] = \int \cL_L
    + \int \bigl \lbrace ^{\rm virtual\ heavy}_{\rm particle\ trees}
    \bigr \rbrace.
}
Thus
\eqn\eHGTREEII{
  \eqalign{\int \delta \cL^{(0)} &= S^{(0)}[\cL_H + \cL_L] -
    S^{(0)}[\cL_L] \cr
    &= \int \bigl \lbrace ^{\rm virtual\ heavy}_{\rm particle\ trees}
    \bigr \rbrace. \cr
  }
}
This is rather trivial, but it is already clear that the matching
correction is a difference between a calculation in the full theory
and one in the low energy effective theory.

We must sharpen the definition of the ``change in the Lagrangian",
$\delta \cL$ (for the moment we consider this object to an arbitrary
number of loops),
in that up to this point it is nonlocal,
because, for example, of its dependence on $p/M$ through the virtual
heavy particle
propagators (here $p$ refers to any of the external momenta).
However, it will turn out to be {\it analytic} in $p/M$ everywhere in
the region
relevant to the low energy theory.  Thus it can be expanded in powers
with the higher order terms steadily decreasing in importance.  It can
then be dealt with, to any finite order in the momentum expansion, as
a local Lagrangian.  We will always interpret the induced operators as
if $\delta \cL$ where expanded to some finite order -- that is, as a
local Lagrangian.  In Feynman diagrams we will indicate the operators
to be momentum expanded by surrounding them by a dotted line; an
example is shown in \fLEBOX, where thick lines represent a
heavy particle and thin lines a light one.  It should be stressed
that the object in \fLEBOX\ indicated by a dotted line is
a point-like, local operator.

\ifx\pansw\pictures
$$
\beginpicture
\setcoordinatesystem units <\tdim,\tdim>
\setplotarea x from -65 to 65 , y from -65 to 30
\stpltsmbl

\plot -20  0  20  0 /
\plot -20 -1  20 -1 /
\plot -20  0  -35 15 /
\plot -20 -1  -35 -16 /
\plot  20  0   35 15 /
\plot  20 -1   35 -16 /
\multiput {.} at -20  10 *10 4 0 /
\multiput {.} at -20 -10 *10 4 0 /
\multiput {.} at -20 -10 *5 0 4 /
\multiput {.} at  20 -10 *5 0 4 /

\put { \baselineskip = 10pt
    \figfont \lines {Figure \xfig\fLEBOX: The low energy approximation\cr
        to a heavy particle propagator (thick line) is indicated\cr
        by surrounding the propagator by a dotted line.\cr}} [t] at 0 -30
\endpicture
$$
\else\fi

Evaluating $\delta \cL$ is more interesting at one loop.  From the
general properties of the effective action, we can write
\eqn\eHGLOOP{
S^{(1)}[\cL_L + \delta \cL] = S^{(1)}[\cL_L + \delta \cL^{(0)}] +
    \int \delta \cL^{(1)}.
}
This is because the one loop term, $\delta \cL^{(1)}$, can appear only
as a tree diagram.  A loop graph with an inclusion of
$\delta \cL^{(1)}$ would give a contribution even higher order in
$\hbar$.  Thus the matching condition to one loop can be
written:
\eqn\eHGMATCH{
  \int \delta \cL^{(1)} = S^{(1)}[\cL_H + \cL_L] -
    S^{(1)}[\cL_L + \delta \cL^{(0)}].
}

In this equation, both $S^{(1)}[\cL_H + \cL_L]$ and
$S^{(1)}[\cL_L + \delta \cL^{(0)}]$ are nonlocal functions of the
fields in which the nonlocality is determined by the long distance
physics.  If there are massless particles in the theory, the 1PI
functions may not even be analytic in the momenta, $p$, as
$p \rightarrow 0$.  However, in the difference, all the long distance
physics that gives rise to this nonanalyticity cancels.  All the
remaining nonlocality is due to the propagation of the heavy particles
and the result can be expanded in powers of $p/M$ to give a sum of
local operators.

The matching calculation is
IR finite.  An infrared divergence always arises from a loop
integration over small momenta.  But for small momentum, the full
theory and the effective theory give the same physics, by
construction.  If we kept an infinite number of terms of the
momentum expansion in the effective theory, the physics would be
exactly the same and there would be no contribution at all from the
loop integral for momenta smaller than $M$.  In practice, to compute the one
loop matching corrections to some order in
the momentum expansion, it is only necessary to include enough terms
to eliminate all the IR divergences in the loop integral for the
matching correction.  If there is still an IR divergence at one loop
to some order in the momentum expansion, it simply means that more terms in
the momentum expansion must be included in the tree level matching
corrections.  This argument works to each order in the loop
expansion. Matching calculations are always completely IR
finite.

We will see that all of the essential features of perturbative matching as
described in this section continue
to hold when we include the effects of instantons.

\newsec{Instantons}

If the functional integral for a relativistic quantum field theory
is defined as an analytic continuation from that of a Euclidean
theory, then solutions and approximate solutions to the Euclidean
equations of motion are
important configurations in the integral.  Expanding
semiclassically about these configurations should in some regime be a good
approximation to the functional integral.  That is all we will say
here about this vast subject.

{}From now on, when we refer to a quantum field theory, we refer to
the Euclidean version.

\subsec{Collective Coordinates}

We briefly review the collective coordinate
formalism \ref\rRAJ{R. Rajaraman, {\it Solitons and Instantons},
(Elsevier Science Publishers, Amsterdam, 1987).}.
Let $S[\Phi]$ be the local classical
action for the bosonic sector of
some quantum field theory, with $\Phi$ being the bosonic
fields. Suppose also that $S$ has some set of continuous
global symmetries.  If $\Phi_{cl}$ is a solution to the equations of
motion, it will break some of these symmetries, and therefore must be
a member of a family of solutions $\Phi_{cl}(a)$, where $a$
represents one parameter for each broken global symmetry degree of
freedom.   We will call these parameters collective coordinates.

Expand the quantum field as $\Phi = \Phi_{cl}(a) + g \eta$, with
$\eta$ the quantum fluctuation, and
$g$ the theory's small coupling.  Then the action can be expanded in
$\eta$:
\eqn\eEXPACT{
    S[\Phi] = S[\Phi_{cl}(a)] +
    {g^2 \over 2} \eta \cdot
    \Biggl( {\delta^2 S \over \delta \Phi^2} \biggr|_{\Phi_{cl}(a)}
    \Biggr)
    \cdot \eta + \cdots
}
Here the dot represents an integral over space-time and a summation
over symmetry/flavor indices.

Define the local operator $\cO(a)$ by
\eqn\eLOCOP{
  {g^2 \over 2} \eta \cdot
  \Biggl( {\delta^2 S \over \delta \Phi^2} \biggr|_{\Phi_{cl}(a)}
  \Biggr)
  \cdot \eta =
  {1 \over 2} \int d^4\/x\ \eta \cO(a,x) \ \eta
}

Let $\zeta_n(a)$ be the orthonormal modes of $\cO(a)$:
\eqn\eORTHMODES{
    \eqalign{
      \cO(a) \zeta_n(a) &= \lambda_n(a) \zeta_n(a) \cr
      \zeta_m(a) \cdot \zeta_n(a) &= \delta_{mn} \cr
    }
}
and expand $\eta = \sum_k \alpha_k \zeta_k(a)$.  The functional
integral can be written:
\eqn\eCCI{
  \eqalign{
    Z &= \int \biggl( \prod_k d \alpha_k \biggr) \e^{-S[\Phi]} \cr
      &= \e^{-S[\Phi_{cl}]} \biggl( \prod_k d \alpha_k \biggr)
         \e^{-1/2 \ \eta \cdot \cO \eta} \cr
  }
}
For each $a$-parameter there will be a zero mode of $\cO(a)$:
\eqn\eZMODE{
    \zeta_k^{(0)}(a) = N_k {\partial \Phi_{cl}(a) \over \partial a_k}
}
where $N_k$ is defined by
\eqn\eZMODENORM{
  \eqalign{
    1 &= \zeta_k^{(0)}(a) \cdot \zeta_k^{(0)}(a) \cr
      &= N_k^2 {\partial \Phi_{cl}(a) \over \partial a_k} \cdot
         {\partial \Phi_{cl} (a) \over \partial a_k} \cr
  }
}
Then $N_k = \bigl({\partial \Phi_{cl}(a) \over \partial a_k} \cdot
             {\partial \Phi_{cl}(a) \over \partial a_k}
\bigr)^{-1/2}$.

As in the Fadeev-Popov method, for each normalizable zero
mode\foot{Only the normalizable zero
modes make the functional determinant zero \ref\rPOLYAKOV{A.M. Polyakov, Nuc.
Phys.
{\bf B120} (1977) 429.}.}\ insert
\eqn\eFP{
  1 = \int da_k\ \delta \bigl(\eta \cdot \zeta^{(0)}_k(a) \bigr)
      {\partial \over \partial a_k}
      \bigl(\eta \cdot \zeta^{(0)}_k(a) \bigr)
}
The $\delta$-function just restricts the fluctuation to be orthogonal
to the zero mode, while
\eqn\eCCBLEAH{
  \eqalign{
    {\partial \over \partial a_k}
    \bigl(\eta \cdot \zeta^{(0)}_k(a) \bigr) &=
    \biggl[{\partial \over \partial a_k}
      \bigl( {\Phi - \Phi_{cl}(a) \over g} \bigr) \biggr]
      \cdot \zeta^{(0)}_k(a) +
      \eta \cdot {\partial \zeta^{(0)}_k(a) \over \partial a_k}
    \cr
    &= - {1 \over g N_k(a)}(1 - Q_k) \cr
  }
}
where
\eqn\eCCINT{
    Q_k = g N_k(a) \; \eta \cdot {\partial
\zeta^{(0)}_k(a) \over \partial a_k}
}
That is, ignoring the overall sign,
\eqn\eCCYUCH{
  1 = {1 \over g} \Biggl(
    \int da_k\ \delta \bigl(\eta \cdot \zeta^{(0)}_k(a) \bigr)
     \bigl({\partial \Phi_{cl}(a) \over \partial a_k} \cdot
             {\partial \Phi_{cl}(a) \over \partial a_k}
             \bigr)^{1/2}
    \Biggr) (1-Q_k)
}

We rederived the well-known
collective coordinate result primarily to emphasize that the
usual formalism can be thought of as a zeroeth order
approximation in $g$: a factor of $Q_k$ comes with one power
of $g$, and usually none is kept.
But since we
will be doing a loop expansion, we must keep all interactions
generated by $Q_k$.  However, when the matching calculation is
performed to any finite order, these interactions will give rise to
only a finite number of diagrams for each term induced in the low
energy theory.

In any case it is no surprise that collective coordinates,
just like semi-classical or perturbative calculations, are not very
meaningful if the coupling is large.

\subsec{Quasi-Collective Coordinates}

In the previous section, we considered collective-coordinates arising from
exact global symmetries.  If we consider the Euclidean action as a height
function over the space of field configurations, these symmetries
appear as exactly level troughs, along the bottom of which lie the
families of exact solutions.  With a collective coordinate, we can perform the
integral along the bottom of the trough exactly.  We
mean to treat families of approximate solutions in a similar way.

For example, suppose $\Phi_{cl}(a)$ and $\Phi_{cl}(a')$ are instantons,
and $a$ and $a'$ are chosen so that the instanton centers $z,z'$ are
widely separated.  Since typically $\Phi_{cl}$ is localized in
space-time, it must be that
$\Phi(a,a') = \Phi_{cl}(a) + \Phi_{cl}(a')$ is a solution to the
equations of motion up to terms which vanish as
$|z-z'|/\rho$ tends to infinity for fixed instanton size $\rho$.

Alternatively, suppose the action is classically scale invariant.  Any
localized nonsingular instanton solution must have a characteristic
size and so break this symmetry.  Then there must be a
collective-coordinate $\mu_I$ corresponding to the (inverse-) size of
the instanton.

If the scale invariance is explicitly broken by adding a
$\Phi$-mass $m$ to the
action, then some aspects of this picture must still hold at scales
much higher than $m$, even though simple scaling arguments imply there
are no exact instanton solutions
\ref\rAFFLECK{I. Affleck, Nuc. Phys. {\bf B191} (1981) 429.}\ref\rTHOOFT{
G. `t Hooft, Phys. Rev. {\bf D14} (3432) 1976.}.
This is because
at high enough scales, the mass
term should be a small perturbation.  So we expect approximate
instantons similar (at distances smaller than $m^{-1}$) to those
present when $m=0$; they will still be naturally characterized by a
scale parameter $\mu_I$, with $\mu_I >> m$.

If we again consider the Euclidean action as a height function over
the space of field configurations, both of these examples appear as
nearly level troughs, along the bottom of which lie families of
approximate solutions parameterized in the first case by $z,z'$ and in
the second by $\mu_I$.  For
$|z-z'|/\rho$ and $\mu_I/m >> 1$ these troughs become
asymptotically flat and should be treated as are the true collective
coordinate troughs to well approximate the functional integral.  Two
ways to do this are the ``constrained instanton" method \rAFFLECK\ and
the ``valley" method
\ref\rVALLEY{
    I.I. Balitsky and A.V. Yung, Phys. Lett. {\bf B168} (1986) 113.}.

In either of these methods, a parameter is introduced for integrating
along the approximately level troughs.  The methods differ essentially
only in that for the valley method the path of this integration always
passes through precisely the lowest point along a slice of field
configurations orthogonal to the trough direction, while the
constrained instanton path passes near, but not necessarily through,
that point.  Since each point of the integration path represents a
classical field configuration (which is then being expanded about) the
definitions of the trough parameters will not be precisely the same,
though both methods should give the same answer for any calculation in
a region where the troughs are nearly flat.
Since there is no need here to
distinguish between these (or any other similar) methods, we will refer
to these trough parameters as quasi-collective coordinates, and to the
classical field configurations as quasi-instantons, without specifying
precisely what this means.

Note that in the first example, the quasi-instanton would be
$\Phi(a,a') = \Phi_{cl}(a) + \Phi_{cl}(a') + \cdots$ where the
unwritten terms are definite, but method dependent functions of
$a$,$a'$.  In the second example the quasi-instanton indexed by
$\mu_I$ is approximately the same as the instanton of size
$\mu_I^{-1}$ from its center out to a radius $r \sim m^{-1}$, after
which it falls off as $\e^{-mr}$ \rAFFLECK.

In the course of doing the quasi-collective coordinate integral,
regions of parameter space will be encountered where quasi-instantons
come closer together than their sizes, and where $\mu_I \sim m$.  Here
the troughs are no longer even approximately flat.  In the first
example will always imagine integrating through these trouble spots
anyway, and in the second integrating $\mu_I$ from some method
dependent scale, of order $m$, up to infinity.  This cavalier approach
is justified as follows: for the first example, we will only be
considering situations where the dilute gas approximation is valid, so
that the functional integral is dominated by field configurations that
look like a collection of widely separated instantons.  If the regions
where instanton overlap are important, than this approximation fails.
For the second example, we will only be considering in detail
situations where the configurations with $\mu_i >> m$ are the
important ones; more specifically there will be a heavy particle mass
$M >> m$ and configurations with $\mu_I << M$ will be unimportant.  We
will show that this is the situation at a heavy quark threshold in
QCD, for example, where the instantons occur in a massless field and
$m=0$.  In section {\bf 4.1}, we will discuss a case where this
condition is not met when the (quasi-) instantons occur in the heavy
field.

Expanding about a quasi-instanton leaves the formalism of the previous
section completely unchanged, except for the effects of expanding
about a different field configuration.  For example, the expansion in
equation
\eEXPACT\ will now have a term linear in the quantum fluctuation.  For
this reason we will typically drop the distinction between instanton
and quasi-instanton, and between collective coordinate and
quasi-collective coordinate, for the rest of this paper.

\subsec{Perturbative Expansion About an Instanton}

Perturbing around an instanton, the generator of $\eta$-correlators is:
\eqn\eETACOR{
    Z[J] = \e^{-S_{cl}} \int (\cD \eta)|_a \;
        {\rm exp} \int \; \bigl[ \cO^{(1)}(a) \eta
            -\eta \cO(a) \eta +
            \cO^{(3)}(a) \eta^3 + \cdots + J \eta \bigr]
}
where $\{ \cO^{(n)}(a) \}$ $(n>2)$ are from the higher order terms
in equation \eEXPACT\ (and $\cO^{(1)}(a)$ is present only in the case
of a quasi-instanton).

Define the instanton density:
\eqn\eDENSE{
  D(a) = \e^{ - S_{cl} }
       \prod_k {1 \over g} \biggl(
       {\partial \Phi_{cl}(a) \over \partial a_k} \cdot
       {\partial \Phi_{cl}(a) \over \partial a_k} \biggr)^{1/2}
}
(where the product is over all zero (or quasi-zero) modes).
Then switching to collective coordinates gives
\eqn\eSWITCH{
  \eqalign{
    Z[J] =&
    \int da \; D(a) \int (\cD \eta)'|_a
    \prod_k (1 - Q_k) \cr
    & {\rm exp} \int \; \bigl[ \cO^{(1)}(a) \eta
        -\eta \cO(a) \eta +
        \cO^{(3)}(a) \eta^3 + \cdots + J \eta \bigr] \cr
  }
}
where $Q_k$ are defined in equation \eCCINT\ and
the prime in $\int (\cD \eta)'|_a$ denotes integration
orthogonal to the zero modes.

Doing the constrained $\eta$-integral is essentially the same problem as
fixing a gauge in standard gauge theory, and we can proceed with a
simplified version of a method often used
there \ref\rPOK{See, e.g., S. Pokorski, {\it Gauge Field Theories},
(Cambridge University Press, Cambridge, 1989).}.
Instead of equation \eFP\ use
\eqn\eNEW{
  1 = \int da_k\ \delta \bigl(\eta \cdot \zeta^{(0)}_k(a) - C_k \bigr)
    {\partial \over \partial a_k}
    \bigl(\eta \cdot \zeta^{(0)}_k(a) \bigr)
}
The rest of the collective coordinate formalism is not modified,
except that now $\eta \cdot \zeta^{(0)}_k(a)$ is restricted to be an
arbitrary constant $C_k$ rather than zero.

Inserting this form of $1$ into the functional integral yields a result
independent of $C_k$, so we can also insert
$\int dC_k\ \e^{-C_k^2/\alpha_k'}$
where $\alpha_k'$ is an arbitrary constant which, like the gauge
fixing parameter $\alpha$, must cancel out at the end of any
calculation of a physical quantity.  This insertion changes the
functional integral by an irrelevant overall constant.
Therefore
\eqn\eTWIDDLE{
  \eqalign{
    Z[J] =& \int da \; D(a) \int (\cD \eta)|_a
    \prod_k (1 - Q_k) \cr
    & {\rm exp} \int \; \bigl[ \cO^{(1)}(a) \eta
        -\eta \widetilde \cO(a) \eta +
        \cO^{(3)}(a) \eta^3 + \cdots + J \eta \bigr] \cr
  }
}
where
\eqn\eMODPROP{
    \eta \cdot \widetilde \cO(a) \cdot \eta =
    \eta \cdot \cO(a) \cdot \eta + \int d^4\/x\ \sum\
    (\eta \cdot \zeta^{(0)}_k)^2/\alpha_k'.
}

For the moment it is most convenient to use the free propagator
$\cO_0$,
treating the rest of the terms as interactions,
because otherwise we aren't certain how to use the formalism of
the second section.  (However, this will be remedied in section
{\bf 3.7}.)  So define
$\cO^{(2)}_1(a) = \widetilde \cO(a) - \cO_0$.  It will also be
convenient to similarly separate the higher order interactions into
those depending on the instanton and those which don't (these last are
$\eta$-self interactions that were present in the original Lagrangian):
$\cO^{(n)}(a) = \cO^{(n)}_0 + \cO^{(n)}_1(a)$.
Then
\eqn\eMORE{
  \eqalign{
    Z[J & ] = \int da \; D(a) \int (\cD \eta)|_a
    \prod_k (1 - Q_k) \cr
    & {\rm exp} \int \; \bigl[ -\eta \cO_0 \eta +
        \cO^{(3)}_0 \eta^3 + \cdots +
        \cO^{(1)}_1(a) \eta +
        \cO^{(2)}_1(a) \eta^2 + \cdots + J \eta \bigr] \cr
  }
}

In \fFEYNMAN\ we introduce a Feynman notation for the theory.

\ifx\pansw\pictures
$$
\beginpicture
\setcoordinatesystem units <\tdim,\tdim>
\setplotarea x from -65 to 65 , y from -80 to 55
\stpltsmbl
\plot -120 40 -90 40 /
\put {=} at -70 40
\put {\figfont $\eta$-propagator} [l] at -50 40

\plot -120 5 -90 5 /
\plot -105 5 -115 -8 /
\plot -105 5 -95 -8 /
\put {\figfont $\cdots$} at -105 -10
\put {\figfont ($n$ $\eta$-legs; $n>2$)} at -105 -20
\put {=} at -70 0
\put {\figfont $\cO_0^{(n)}$} [l] at -50 0

\put {$\bullet$} at 55 40
\put {=} at 90 40
\put {\figfont $\int da\ D(a)$} [l] at 110 40
\put {$\bullet$} at 55 20
\multiput {\phdr} at 55 20 *1 0 -10 /
\plot 40 0 70 0 /
\plot 55 0 45 -13 /
\plot 55 0 65 -13 /
\put {\figfont $\cdots$} at 55 -15
\put {\figfont ($n$ $\eta$-legs; $n>0$)} at 55 -25
\put {=} at 90 0
\put {\figfont $\int da\ D(a)\ \cO^{(n)}_1(a)$} [l] at 110 0
\put {\figfont Figure \xfig\fFEYNMAN: Feynman notation.} at 0 -65
\endpicture
$$
\else\fi

\subsec{Dilute Gas Approximation}

The basic assumption of the dilute gas approximation (DGA) is that the
functional
integral is dominated by field configurations that look like a
collection of widely separated instantons.  Corrections can be added
order by order in instanton density $D$.

In the $n$-instanton sector the action can be expanded as
\eqn\eEXPS{
  \eqalign{
    S \biggl[\sum_{I=1}^n \Phi_{cl} & (a_I) + \eta \biggr] = n S_{cl} +
        \int \; \bigl[ -\eta \cO_0 \eta +
            \cO^{(3)}_0 \eta^3 + \cdots \bigr] \cr
        & + \sum_I \int \bigl[ \cO^{(1)}_1(a_I) \eta + \cdots \bigr] \cr
        & \;\; + \sum_{k=2}^n \int (k-{\rm body\ \;\; instanton-instanton\
            \;\; interactions)} \cr
  }
}
The sum over instanton-instanton interactions is defined so as to make
the equation correct; a $k$-body
term consists of a function of $k$ instanton
coordinates coupled to any number (including zero) of $\eta$'s.
In principle the sum could be computed, but we will see that it
contributes at next to leading order in $D$.
The $\{ \cO^{(n)}_1 \}$ are just the
1-body interactions written out more explicitly.

The generator of $\eta$-correlators is now
\eqn\eNETACOR{
  \eqalign{
    Z_{\rm DGA}[J & ] = \sum_{n=0}^\infty {1 \over n!}
        \biggl( \prod_{I=1}^n \int da_I\ D(a_I) \biggr) \int \cD \eta
        \prod_{I=1}^n \biggl[ \prod_k \bigl( 1 - Q_k(a_I) \bigr)
            \biggr] \cr
    & {\rm exp} \Bigl \lbrace \int \bigl[ -\eta \cO_0 \eta +
            \cO^{(3)}_0 \eta^3 + \cdots \bigr] +
        \sum_{I=1}^n \int \bigl[ \cO^{(1)}(a_I) \eta + \cdots \bigr]
        + \int J \eta \cr
    & \;\; + \sum_{k=2}^n \int (k-{\rm body\ \;\; instanton-instanton\ \;\;
            interactions)}  \Bigr \rbrace \cr
  }
}
(The factor of $n!$ is necessary to avoid double counting identical
instantons.)
Here the sum over $k$-body interactions, which has the same basic form as
before, accounts not only for the errors in expanding the action,
but also in the collective coordinate procedure.  Since we have
assumed the same zero modes etc. as if the instantons were infinitely
separated, this procedure requires corrections that look like
instanton-instanton interactions etc.  Again, the sum could
in principle
be computed, but will contribute at next to leading order in $D$.

Absorb
the factors of $1-Q_k$ into the interactions by using
$1-Q = \e^{{\rm ln}(1-Q)}$; this amounts to setting
\eqn\eABSORB{
    \cO^{(n)}_1(a) \rightarrow \cO^{(n)}_1(a) -
        {1 \over n} \sum_k Q_k^n(a)
}
It follows that
\eqn\eEXPDGA{
  \eqalign{
    Z_{\rm DGA}[J] = \int \cD \eta\; & {\rm exp} \Bigl \lbrace
        \int \bigl[ - \eta \cO_0 \eta + \cO_0^{(3)} \eta^3 + \cdots +
            J \eta \bigr] + \cr
        & \; \int da\  D(a)\ \e^{\int \bigl[ \cO^{(1)}_1(a) \eta +
            \cO^{(2)}_1(a) \eta^2 + \cdots \bigr]} + \cO(D^2)
    \Bigr \rbrace \cr
  }
}
We are dealing explicitly with only leading order in $D$, but it
should be stressed that we can always calculate to higher order.

Now, in addition to the diagrams shown in \fFEYNMAN, there are
interactions at $\cO(D^1)$
obtained from products of 1-body operators; an example
is shown in \fPROD.
\ifx\pansw\pictures
$$
\beginpicture
\setcoordinatesystem units <\tdim,\tdim>
\setplotarea x from -65 to 65 , y from -50 to 45
\stpltsmbl
\plot -15 20 15 20 /
\plot 0 20 -10 33 /
\plot 0 20 10 33 /
\multiput {\phdr} at 0 20 *3 0 -10 /
\plot -15 -20 15 -20 /
\plot -35 0 -20 0 /
\multiput {\phru} at -10 0 *1 -10 0 /
\put {$\bullet$} at 0 0
\put { \baselineskip = 10pt
    \figfont \lines {Figure \xfig\fPROD: Diagram for the \cr
    interaction
    $\int da\ D(a)\ [\cO^{(1)}_1(a) \eta]
    [\cO^{(4)}_1(a) \eta^4] [\cO^{(2)}_1(a) \eta^2]$.\cr}} at 0 -35
\endpicture
$$
\else\fi

\subsec{DGA as a map from Lagrangians to Lagrangians}

It will be convenient to think of the DGA as a {\it map $F$ from
Lagrangians to Lagrangians}.  That is, if $\cL$ is the Lagrangian
before including instanton effects, so that the generator of
$\Phi$-correlators is
\eqn\ePHICOR{
    Z[J] = \int \cD \Phi \; \e^{\int \bigl( -\cL + J \Phi \bigr)}
}
then the DGA generator of $\eta$-correlators can be written as
\eqn\eDGAETA{
    Z_{\rm DGA}[J]
        = \int \cD \eta \; \e^{\int \bigl( -F[\cL] + J \eta \bigr)}
}
where the functional integral is to be performed perturbatively in the
trivial instanton sector.
{}From \eEXPDGA\ it follows that
\eqn\eFSTUFF{
    \int F[\cL] = \int \cL - \int da\ D(a)\
        \e^{\int [ \cO^{(1)}_1(a) \eta + \cdots ]} + \cO(D^2)
}

The term ``Lagrangian" has been used loosely here as in \eHGTREEII\ and
\eHGMATCH.  If $\cL$ is a
local Lagrangian, then generally $F[\cL]$ will be nonlocal because of
the nonlocality of the instanton induced interactions.  However we
will see that in a matching calculation this nonlocality is completely
under control, in that the matching corrections, when calculated to any
finite order in the momentum expansion, will be local.  But any instantons
remaining in the effective theory will still generally give nonlocal
interactions.

\subsec{Renormalization and Renormalization Group}

Wherever we write the instanton density $D$ or coupling $g$ we mean
the renormalized instanton density $D^{(\mu)}$ and coupling $g(\mu)$,
where $\mu$ is the renormalization group scale.  $D^{(\mu)}$ is
defined by \eDENSE, where any couplings occurring in the right
hand side are the renormalized ones.
As explained in the introduction, a mass independent scheme is
desirable so we imagine choosing the counterterms via MS.

\subsec{Nonperturbative interactions vs propagating in an instanton
background}

There is no small coupling associated with an instanton interaction
except for the instanton density.  A diagram where the same instanton
interacts several times with a given quantum particle line (with no
other instanton interactions) is still $\cO(D^1)$.  Therefore,
whenever we are considering a diagram with one or more instantons, we
must imagine summing up all diagrams where each instanton interacts
with each quantum line any number of times.  A notation for this is
indicated in \fSUMEM.
\ifx\pansw\pictures
$$
\beginpicture
\setcoordinatesystem units <\tdim,\tdim>
\setplotarea x from -65 to 65 , y from -45 to 55
\stpltsmbl
\put {$\bullet$} at -115 0
\multiput {\beginpicture
    \multiput {\phdr} at 0 20 *1 0 -10 /
    \plot -15 20 15 20 /
    \plot 0 20 -10 33 /
    \plot 0 20 10 33 /
\endpicture} [b] at -115 10    -65 10    0 10 /
\multiput {\jagdr} at -115 0 *1 0 -10 /
\plot -130 -20 -100 -20 /
\put {$=$} at -90 5
\put {$\bullet$} at -65 0
\multiput {\phdr} at -65 0 *1 0 -10 /
\plot -80 -20 -50 -20 /
\put {$+$} at -40 5
\put {$\bullet$} at 0 0
\startrotation by 0.66667 -0.74536 about 0 0
    \put {\phru} at 0 0
    \multiput {\phrd} at 5 0 *1 10 0 /
\stoprotation
\startrotation by -0.66667 -0.74536 about 0 0
    \put {\phrd} at 0 0
    \multiput {\phru} at 5 0 *1 10 0 /
\stoprotation
\plot -32 -19   32 -19 /
\put {$+\;\;\;\cdots$} at 47 5
\put {\figfont Figure \xfig\fSUMEM: The jagged line indicates
    the sum on the right.}  at -35 -30
\endpicture
$$
\else\fi
Actually, the sum indicated by the jagged line
should be thought of as the particle propagator in an instanton
background (before integrating over the instanton's collective
coordinates),
minus the free propagator, rather than as a series of
nonperturbative interactions.

However, when an instanton connects two different quantum lines, or
interacts with the same line both before and after another instanton
does, the diagram should be thought of as having a nonperturbative
interaction.  For the purposes of counting loops etc, these
interactions can be shrunk to a point.  An example is shown in
\fSHRINKEM.
\ifx\pansw\pictures
$$
\beginpicture
\setcoordinatesystem units <\tdim,\tdim>
\setplotarea x from -65 to 65 , y from -75 to 15
\stpltsmbl
\put {$\bullet$} at 0 0
\startrotation by 0.66667 -0.74536 about 0 0
    \put {\jagru} at 0 0
    \multiput {\jagrd} at 5 0 *1 10 0 /
\stoprotation
\startrotation by -0.66667 -0.74536 about 0 0
    \put {\jagrd} at 0 0
    \multiput {\jagru} at 5 0 *1 10 0 /
\stoprotation
\plot -32 -19   32 -19 /
\multiput {\jagdr} at 0 -19 *1 0 -10 /
\put{$\bullet$} at 0 -39
\put {$\rightarrow$} at 42 -14
\plot 52 -10  82 -10 /
\put{$\bullet$} at 67 -10
\circulararc 360 degrees from 77 -20 center at 67 -20
\put{$\bullet$} at 67 -30
\put { \baselineskip = 10pt
\figfont \lines {
   Figure \xfig\fSHRINKEM: For counting loops, nonperturbative interactions
   can be\cr
   shrunken to a point.\cr}}
       at 27 -50
\endpicture
$$
\else\fi

\newsec{Matching in Effective Field Theories With Instantons}

As pointed out in \rHGREV,
there are two cases to consider.

\subsec{Instantons in the Heavy Field}

Let $\Phi$ represent now just the light fields, and suppose $\Psi$ is
a heavy bosonic field, of mass $M$, with instantons.  As explained in
section {\bf 3.2}, the instantons will be indexed by a
quasi-collective coordinate $\mu_I$ (and have characteristic size
$\mu_I^{-1}$) where $\mu_I$ runs from some scale above $M$ to
infinity.  Most of an instanton's Fourier transform will be in momenta
higher than $\mu_I$, and in particular, higher than $M$.  The
low energy theory has no $\Psi$ and therefore no instantons; they will
appear only through the matching corrections.  The fact that the
instanton is cutoff at frequencies below $M$ means that the momentum
expansion in the low energy theory will be good.

The matching equation can be taken to be
\eqn\ePERTMATCH{
    S_{\rm pert} \bigl[ F[ \cL_H + \cL_L ] \bigr] =
        S_{\rm pert}[ \cL_L + \delta \cL ]
}
where $S_{\rm pert}$ is the 1LPI effective action, evaluated
perturbatively about the trivial vacuum,
and the renormalization scale is
$\mu = M$.  Note that $S_{\rm pert}[\cL]$ is defined to be the
functional
constructed via a loop expansion from the interactions present
in $\cL$, whether or not some of those interactions arise from
instantons.  In particular, suppose $\cL$ is the Lagrangian for some
theory with instantons (although no instanton effects on $\cL$ have
yet been included), and let $S[\cL]$ be the theory's effective action.
Then even assuming the validity of the loop expansion, we do not
expect $S_{\rm pert}[\cL]=S[\cL]$ since $S_{\rm pert}$ ignores
instanton
effects; rather we expect (within the loop and dilute gas
approximations) that $S_{\rm pert} \bigl[ F[\cL] \bigr] = S[\cL]$.

Because the instantons occur in the heavy field, the quasi-collective
coordinate method may fail.  This is because instantons with
$\mu_I$-values ranging all the way down to their
quasi-instanton-scheme dependent cutoff scale $\sim M$ may be
important.  If the calculation is scheme dependent, then a better
method for dealing with the quasi-instantons is needed to calculate
the matching corrections exactly.  However, since all schemes should
give roughly similar answers, the matching corrections can at least be
estimated.

\subsec{Instantons in a Light Field}

Now suppose one of the light bosonic fields, $\phi$, has instantons,
and let $\Psi$ be a heavy bosonic or fermionic field of mass $M$.
This situation is more complicated because there will generally be
instanton effects both in the full and effective theories.  At scales
far above the $\phi$-mass (or at any scale if $\phi$ is massless) the
$\phi$-sector of the action will be approximately scale invariant, and
therefore the instantons must come in families indexed by size.  Then,
as explained in the third section, there will be a collective
coordinate $\mu_I$ corresponding to the instanton scale; $\mu_I$ must
be integrated from some scale above the $\phi$-mass to infinity.  For
a given $\mu_I$ the instanton's Fourier transform will be cutoff by
the $\phi$-mass in the IR and by $\mu_I$ in the UV.

If the $\phi$ is massless, as in QCD, then instanton calculations in
the theory may suffer severe IR problems from the
$\mu_I \rightarrow 0$ limit.  Also, as explained in the introduction,
small instantons ($\mu_I > M$) would be problematic if present in the
low energy theory because of their incompatibility with the
low energy approximations to operators induced by $\Psi$.

The matching calculation is free of the first problem because of the
cancellation of IR physics, and naturally
avoids the second one.  The low energy theory is still affected by
instantons both with $\mu_I$ above and below $M$.  But the small ones
($\mu_I > M$) are present only in the matching corrections.

To realize this, it will be convenient to
generalize the DGA map $F$ as follows: Let $F_\Lambda[\cL]$ be
the Lagrangian obtained by applying the DGA to $\cL$, using the
$\phi$-instantons, but only integrating $\mu_I$ up to the scale
$\Lambda$.  (If $\Lambda$ is below the scale where $\phi$-instantons
exist, then $F_\Lambda[\cL] = \cL$.)

The matching condition, applied at $\mu=M$, can be taken to be
\eqn\eMATCH{
    S_{\rm pert} \bigl[ F_\infty [ \cL_H + \cL_L ] \bigr]
    = S_{\rm pert} \bigl[ F_M [ \cL_L + \delta \cL ] \bigr]
}
where $S_{\rm pert}$ is the 1LPI effective action,
evaluated perturbatively about the trivial vacuum.
Any arbitrariness on the right hand side is just a definition of
$\delta \cL$.

To solve for $\delta \cL$,
expand in loops and renormalized instanton density $D^{(\mu)}$
(here $\cL$ is an arbitrary Lagrangian):
\eqn\eEXPDL{
  \eqalign{
    \cL &= \sum_{l,d} \cL^{(l,d)} \cr
    S_{\rm pert}[\cL] &= \sum_{l,d} S_{\rm pert}^{(l,d)}[\cL] \cr
    \delta \cL &= \sum_{l,d} \delta \cL^{(l,d)} \cr
    F_\Lambda[\cL] &= \sum_{l,d} F_\Lambda^{(l,d)}[\cL] \cr
  }
}
It should be stressed that $S_{\rm pert}^{(l,d)}[\cL]$ and
$F_\Lambda^{(l,d)}[\cL]$ are projections of $S_{\rm pert}$ and $F_\Lambda$
onto
the $l$-loop, $d$-instanton density part.  If there are terms from
loop effects or involving factors of instanton density already present
in $\cL$, then the $(l,d)$-order of these terms must be taken into
account when calculating $S_{\rm pert}^{(l,d)}[\cL]$ and
$F_\Lambda^{(l,d)}[\cL]$.

It is now possible to solve for $\delta \cL$ order
by order in $(l,d)$.
We define the following partial sums:
\eqn\eSUMNOT{
  \eqalign{
        G^{[l,d]} &= \sum^l_{l'=0} \sum^d_{d'=0}
        G^{(l',d')}\cr
        G^{[l,d)} &= \sum^l_{l'=0}
        G^{(l',d)}\cr
        G^{(l,d]} &= \sum^d_{d'=0}
        G^{(l,d')}\,,
  }
}
where $G$ is any of the objects $\cL$, $\delta\cL$, or $F_\Lambda$.
As explained in the second section, by cluster arguments:
\eqn\eCLUSTER{
    S_{\rm pert}^{(l,d)}[\cL^{[l-1,d]} + \cL^{(l,d]}] =
    S_{\rm pert}^{(l,d)}[\cL^{[l-1,d]}] + \int \cL^{(l,d)}.
}
By counting powers of $D^{(\mu)}$, there is a similar rule for $F_{\Lambda}$:
\eqn\ePOWD{
    F_{\Lambda}^{(l,d)}[ \cL^{[l,d-1]} + \cL^{[l,d)}] =
    F_{\Lambda}^{(l,d)}[ \cL^{[l,d-1]} ] + \cL^{(l,d)}
}
Note that \ePOWD\ implies
\eqn\eFPROP{
    F_\Lambda^{(l,0)}[\cL^{(l,0)}] = \cL^{(l,0)}
}
With these rules, if we know $\delta\cL$ to
$\cO(l-1,d)$ and to $\cO(l,d-1)$, we can compute it to
compute to $\cO(l,d)$ as follows:
\eqn\eDERIV{
  \eqalign{
    &\vbox{\hsize70ex $ S_{\rm pert}^{(l,d)} \bigl[ F_\infty [\cL_H +
    \cL_L] \bigr] = S_{\rm pert} \bigl[ F_M [ \cL_L + \delta \cL ] \bigr]$}\cr
    &\vbox{\hsize70ex $= S_{\rm pert}^{[l,d]} \bigl[ F_M^{[l,d]}[\cL_L +
\delta\cL^{[l,d]}]
\bigr]
$}\cr
    &\vbox{\hsize70ex $= S_{\rm pert}^{(l,d)} \bigl[ F_M^{[l,d]}[\cL_L +
\delta\cL^{[l,d-1]}]
        + \delta \cL^{[l,d)}  \bigr]$
        \hfill (by eq{.} \ePOWD)} \cr
    &\vbox{\hsize70ex $= S_{\rm pert}^{(l,d)} \bigl[ F_M^{[l-1,d]}[\cL_L +
\delta\cL^{[l-1,d-1]}]
        + \delta \cL^{[l-1,d)}  $}\cr
        &\vbox{\hsize70ex $ \;\;\;\;\;\;\;\;\;
        + F_M^{(l,d]}[\cL_L + \delta\cL^{[l,d-1]}] + \delta \cL^{(l,d)}
\bigr]$
        \hfill (by def{.} of $F$)} \cr
    &\vbox{\hsize70ex $= S_{\rm pert}^{(l,d)} \bigl[ F_M^{[l-1,d]}[\cL_L +
\delta\cL^{[l-1,d-1]}]
        + \delta \cL^{[l-1,d)}  \bigr] $}\cr
        &\vbox{\hsize70ex $ \;\;\;\;\;\;\;\;
        + \int \bigl[ F_M^{(l,d)}[\cL_L + \delta\cL^{[l,d-1]}] +
            \delta \cL^{(l,d)} \bigr]$
        \hfill (by eq{.} \eCLUSTER)} \cr
  }
}
Thus
\eqn\eGENSOL{
  \eqalign{
    \int \delta \cL^{(l,d)} &= S_{\rm pert}^{(l,d)} \bigl[ F_\infty [\cL_H +
    \cL_L] \bigr] \cr
    &- S_{\rm pert}^{(l,d)} \biggl[ F_M^{[l-1,d]} [\cL_L +
    \delta\cL^{[l-1,d-1]} ] + \delta \cL^{[l-1,d)}  \biggr]\cr
    &- \int F_M^{(l,d)} [\cL_L + \delta\cL^{[l,d-1]}] \cr
  }
}
Equation \eGENSOL\ is the principle result of this paper. It is the
generalization of the matching relations of \rHGREV\ to include instanton-like
effects. As expected it depends on $\delta\cL$ up to $\cO(l-1,d)$ and to
$\cO(l,d-1)$. Because it is still a matching correction, we also expect that
the nonperturbative $\delta\cL$ is IR finite and is nonlocal only on the heavy
particle scale. We will show expicitly how this works in low orders in the
examples discussed below.

\subsec{Anti-instantons}

We have been referring to all time-dependent non-trivial solutions of
the Euclidean equations of motion as ``instantons", but if the bosonic
action is parity invariant then for any solution there will always be
a parity reversed solution. If the two solutions are different then
convention determines which is called the instanton and which the
anti-instanton.

Precisely the same factors of instanton density are associated with a
diagram containing anti-instanton interactions as would be were there
only instantons, and for each leading order instanton-interaction
there is precisely the same anti-instanton-interaction.  Therefore the
previous formalism is essentially unchanged.  Instantons and
anti-instantons are distinguishable from one another, however, so we
introduce a Feynman notation in \fANTI; \fANTI\ is to be
compared with \fFEYNMAN.

\ifx\pansw\pictures
$$
\beginpicture
\setcoordinatesystem units <\tdim,\tdim>
\setplotarea x from -65 to 65 , y from -80 to 55
\stpltsmbl
\put {$\circ$} at -35 40
\put {=} at 0 40
\put {\figfont $\int da\ D(a)$} [l] at 20 40
\put {$\circ$} at -35 20
\multiput {\phdr} at -35 20 *1 0 -10 /
\plot -50 0 -20 0 /
\plot -35 0 -45 -13 /
\plot -35 0 -25 -13 /
\put {$\cdots$} at -35 -15
\put {\figfont ($n$ $\eta$-legs)} at -35 -25
\put {=} at 0 0
\put {\figfont $\int da\ D(a)\ \overline \cO^{(n)}_1(a)$} [l] at 20 0
\put { \baselineskip = 10pt \figfont \lines {
    Figure \xfig\fANTI: Feynman notation for anti-instantons. \cr
    Here $\overline \cO^{(n)}_1(a)$ is the anti-instanton \cr
    version of $\cO^{(n)}_1(a)$ (see \fFEYNMAN). \cr}} at 0 -65
\endpicture
$$
\else\fi

\newsec{QCD}

As an example, we imagine matching out a heavy quark, mass $M$, in
QCD, to order (1,1).  We assume that the DGA is good at the scale
set by $M$ for instantons with
$\mu_I > M$.

After gauge fixing, the Lagrangian is
\eqn\eQCDLAG{
    \cL = - {1 \over 2 g^2} Tr\ F^2(\cA)
    + {1 \over 2 \alpha} Tr\ G^2(\cA) +
    \sum_F\ \overline \psi_F (i\spur D - m_F)\psi_F +
    \Phi^\dagger {\delta G \over \delta \Omega} \Phi
}
$F_{\mu \nu}$ and $D_\mu$ are, respectively, the field strength and
covariant derivative for the gauge field $\cA$, and  $\{ \psi_F \}$
are the fermions. The gauge group $\cG$ is $SU(3)$ but we will act as
if it were
$SU(2)$ without loss of
generality, since for a larger gauge group the instantons will live in
the $SU(2)$ subgroups \ref\rCOLEMAN{S. Coleman, {\it Aspects of
Symmetry}, (Cambridge University Press, Cambridge, 1988).}.
(From now on, we will suppress the
fermion flavor index $F$.)  $G[\cA]$ is a gauge fixing
functional, and $\Phi$ is the fermionic complex scalar Fadeev-Popov
ghost; ${\delta G \over \delta \Omega}$ is the variation of $G$ with
respect to an infinitesimal gauge transformation.  Finally, $g$ and
$\alpha$ are, respectively, the gauge coupling and gauge fixing
parameters.

In singular gauge the instanton is \ref\rANDGROSS{See, e.g., N. Andrei
and D.J. Gross, Phys. Rev. {\bf D18} (1978) 468.}
\eqn\eINST{
    A^a_\mu(x;R,\mu,z) = {R_{ab} \eta_{b \mu \nu} (x - z)_\nu \over
    (x - z)^2 [1 + \mu^2(x-z)^2]}
}
where $R$ is an $SU(2)$ rotation matrix corresponding to the instanton's
group orientation, $\mu$ is the instanton's inverse size, and $z$ its
position.  The $\eta$-symbol is defined in \rTHOOFT; it maps
antisymmetric tensors $T_{\mu \nu}$ in $SO(4)$, which is locally
equivalent to $SO(3) \times SO(3)$, onto 3-vectors in one of the $SO(3)$
groups.  The anti-instanton $\overline A$ has the same expression but
with $\eta \rightarrow \overline \eta$, where $\overline \eta$ maps
into the other $SO(3)$.

We can define a multi-instanton configuration as
\eqn\eMULTII{
    ``A_{n,\overline n}" = \sum^n_{i=1} A(z_i,R_i,\mu_i) +
    \sum^{\overline n}_{i=1} {\overline A}(z_i,R_i,\mu_i)
}
We place $``A_{n,\overline n}"$ in quotes because we actually define
$A_{n,\overline n}$ to be the above expression after a singular gauge
transformation such that it is no longer singular.

\subsec{Background Field Gauge}

The expansion about an
instanton necessitates partial use of the background field method
\ref\rABBOTT{L.F. Abbott, Nuc. Phys. {\bf B185} (1981) 189}\rPOK.
We write $\cA = A + g \eta$, where $A$ is a background
classical field configuration and $\eta$ a quantum fluctuation.

Any physical quantity we calculate will be invariant under ``quantum"
gauge transformations:

\eqn\eQUANTUM{
    \eqalign{
      A & \rightarrow\ A \cr
        \eta\ & \rightarrow\ {\bf g} \eta {\bf g}^{-1} +
            {\bf g}\;d{\bf g}^{-1} + {\bf g}\;A\;{\bf g}^{-1} - A \cr
    }
}
(with ${\bf g}$ any $\cG$-valued function).
We can presumably always fix $\eta$, via a quantum gauge
transformation, to background field gauge:
$d \eta + [A,\eta] = 0$

With this choice of gauge, physical quantities will be invariant under
``background" gauge transformations:
\eqn\eBACKGROUND{
    \eqalign{
      A\ & \rightarrow\
            {\bf g}\;A\;{\bf g}^{-1} + {\bf g}\;d{\bf g}^{-1} \cr
      \eta\ & \rightarrow\ {\bf g} \eta {\bf g}^{-1} \cr
    }
}

Much of the utility of the background field method is that it calculates
the effective action while avoiding calculating (and
renormalizing) diagrams with external $\eta$-legs, but we won't take
advantage of this feature, since the $\eta$-field will be kept as the gluon
degree of freedom, and we are not concerned with maintaining explicit
gauge invariance.

\subsec{Instanton Density}

Equations \eDENSE\ and \eINST\ give
\eqn\eQCDD{
    D(a)\;da = ({\rm const})
    g^{\rm -(\#\ zero\ modes)}\ \e^{- 8 \pi^2 / g^2}\;
    dR\;\mu^3 d \mu\; d^4\/z
}
where $R$, $\mu$, $z$ are as in equation \eINST\ and $dR$ is the
group's invariant measure; both the number of zero modes and the
overall constant depend on the gauge group $\cG$.

\subsec{Feynman Notation}

We adopt a notation essentially the same as before.  The propogators
are shown in \fQCDFEYN, and a subset of the interactions at $\cO(D^1)$
are indicated in \fQCDINTS.

\ifx\pansw\pictures
$$
\beginpicture
\setcoordinatesystem units <\tdim,\tdim>
\setplotarea x from -65 to 65 , y from -95 to 15
\stpltsmbl
\multiput {\sru}  at -50 0 *4 10 0 /
\put {\figfont $= \;\;\;\eta$} [l] at 10 0
\plot -50 -20   0 -20 /
\tarrow from -27 -20 to -22 -20
\put {\figfont $= \;\;\;$light fermion} [l] at 10 -20
\put {\beginpicture
    \setplotsymbol ({\tenrm .})
    \plot 0 0 50 0 /
    \plot 0 -1  50 -1 /
    \endpicture} at -50 -40
\tarrow from -27 -40 to -22 -40
\tarrow from -27 -41 to -22 -41
\put {\figfont $= \;\;\;$heavy fermion} [l] at 10 -40
\multiput {\plot 0 0 6 0 / }  at -50 -60 *4  11 0 /
\tarrow from -27 -60 to -22 -60
\put {\figfont $= \;\;\;$ghost} [l] at 10 -60
\put {\figfont Figure \xfig\fQCDFEYN: Feynman notation for QCD propagators.}
    at -10 -80
\endpicture
$$
\else\fi

The complete set of interactions up to
this order in $D$ involves all the usual QCD interactions, and
all interactions obtained by connecting instantons in \fQCDINTS\
(an example of this was given in \fPROD).  Note that there are
interactions, coming from the collective coordinate terms, with an
instanton connected to an arbitrary number of gluons.

\ifx\pansw\pictures
$$
\beginpicture
\setcoordinatesystem units <\tdim,\tdim>
\setplotarea x from -65 to 65 , y from -110 to 15
\stpltsmbl
\plot -95 -20   -45 -20 /
\put {\beginpicture
    \setplotsymbol ({\tenrm .})
    \plot 0 0 50 0 /
    \plot 0 -1  50 -1 /
    \endpicture} at -25 -20
\multiput {\plot 0 0 6 0 / }  at 45 -20 *4  11 0 /
\multiput {\beginpicture
    \put {$\bullet$} at 0 0
    \multiput {\phdr} at 0 0 *1 0 -10 /
    \endpicture} at -70 -3    0 -3    70 -3 /
\put {$\bullet$} at -95 -55
\multiput {\phru} at -95 -55 *1 10 0 /
\multiput {\sru}  at -75 -55 *1 10 0 /
\put {$\bullet$} at -25 -55
\multiput {\phru} at -25 -55 *1 10 0 /
\multiput {\sdr} at -5 -35 *3 0 -10 /
\put {$\cdots$} at 70 -55
\put { \baselineskip = 10pt \figfont \lines {
    Figure \xfig\fQCDINTS: Subset of $\cO(D^1)$ interactions. \cr
    For each instanton interaction, there is \cr
    a corresponding anti-instanton interaction. \cr}} at 0 -95
\endpicture
$$
\else\fi

\subsec{QCD Calculation to order (0,0)}

To this order, equation \eGENSOL\ gives
\eqn\eMATCHOHOH{
    \int \delta \cL^{(0,0)} =
        S_{\rm pert}^{(0,0)} \bigl[ F_\infty^{(0,0)} [\cL_H + \cL_L] \bigr]
        - \int F_M^{(0,0)}[\cL_L]
}
Using \eFPROP\ and the fact that there are no virtual heavy particle trees in
QCD, gives
$\int \delta \cL^{(0,0)} = 0$.

\subsec{QCD Calculation to order (1,0)}

To this order, equation \eGENSOL\ gives
\eqn\eMATCHWONOH{
  \eqalign{
    \int \delta \cL^{(1,0)} &= S_{\rm pert}^{(1,0)}
    \bigl[ F_\infty [\cL_H + \cL_L] \bigr] -
    S_{\rm pert}^{(1,0)} \bigl[ F_M^{[0,0]}[\cL_L] +
    \delta \cL^{[0,0)}  \bigr] \cr
    & \;\;\;\;\;\;\; - \int F_M^{(1,0)}[\cL_L] \cr
    &=
    S_{\rm pert}^{(1,0)} \bigl[ \cL_H + \cL_L \bigr]- S_{\rm pert}^{(1,0)}
\bigl[ \cL_L \bigr]
    \cr
  }
}
On the right hand side, all the diagrams cancel except the heavy fermion
loops.
That is, $\int \delta \cL^{(1,0)}$
is the sum over
(the low energy
approximations to) heavy fermion loops, with
gluon insertions (see \fDLONEZERO).

\ifx\pansw\pictures
$$
\beginpicture
\setcoordinatesystem units <\tdim,\tdim>
\setplotarea x from -60 to 60 , y from -45 to 70
\stpltsmbl
\put {\figfont $\int \delta \cL^{(1,0)} =$} [r] at -60 0
\multiput {\hfloop} at 0 0   115 0 /
\multiput {\lebox} at -1 0   114 0 /
\multiput {\beginpicture
               \multiput {\sru}  at 0 0 *2 10 0 /
           \endpicture} at -35 0   25 0   80 0    140 0 /
\multiput {\sdr} at 115 25 *2 0 10 /
\put {$+$} at 58 0
\put {$+\;\;\cdots$} [l] at 170 0
\put {\figfont Figure \xfig\fDLONEZERO: Matching contribution
    at (1,0).} [l] at -30 -30
\endpicture
$$
\else\fi

\subsec{QCD Calculation to order (0,1)}

To this order, equation \eGENSOL\ gives
\eqn\eMATCHOHWON{
  \eqalign{
    \int \delta \cL^{(0,1)} &=
    S_{\rm pert}^{(0,1)} \bigl[ F_\infty [\cL_H + \cL_L] \bigr] -
    \int F_M^{(0,1)} [\cL_L] \cr
    &= \int \bigl( F_\infty^{(0,1)}[\cL_L] - F_M^{(0,1)}[\cL_L] \bigr) \cr
  }
}
On the right hand side occur all the instanton-light particle interactions,
and there is a cancellation for all instantons with $\mu_I < M$.
That is, $\int \delta \cL^{(0,1)}$ is the sum over (the low energy
approximation to) all the instanton - light particle
interactions, with the instanton scales all running from $M$ to $\infty$
(see \fDLZEROONE).

\ifx\pansw\pictures
$$
\beginpicture
\setcoordinatesystem units <\tdim,\tdim>
\setplotarea x from -60 to 40 , y from -55 to 35
\stpltsmbl
\put {\figfont $\int \delta \cL^{(0,1)} =$} [r] at -30 0
\multiput {\jagdr}  at 0 20 *3 0 -10 /
\multiput {\srd}  at -20 -20 *5 10 0 /
\plot -20 20 40 20 /
\put {$\bullet$} at 0 0
\put {$> M$} at 15 0
\put {$+\;\;\cdots$} [l] at 55 0
\multiput {.} at -10 -35 *16 0 4 /
\multiput {.} at -10 33 *10 4 0 /
\multiput {.} at 30 -35 *16 0 4 /
\multiput {.} at -10 -35 *10 4 0 /
\put { \baselineskip = 10pt \figfont \lines {
    Figure \xfig\fDLZEROONE: Matching contribution at (0,1). \cr
    For each instanton diagram, there is \cr
    a corresponding anti-instanton diagram. \cr}} at 5 -60
\endpicture
$$
\else\fi

\subsec{QCD Calculation to order (1,1)}

To this order equation
\eGENSOL\ gives
\eqna\eIMPEN
$$
  \eqalignno{
    \delta \cL^{(1,1)} &=
    S_{\rm pert}^{(1,1)} \bigl[ F_\infty [\cL_H + \cL_L] \bigr] -
    S_{\rm pert}^{(1,1)} \bigl[ F_M^{[0,1]}[\cL_L] +
    \delta \cL^{[0,1)}  \bigr] \cr
    & \;\;\;\;\;\;\;\;\; - \int F_M^{(1,1)} [\cL_L + \delta\cL^{[1,0]}] \cr
    &= S_{\rm pert}^{(1,1)} \bigl[ \cL_H + \cL_L +
    F_\infty^{(0,1)} [\cL_H + \cL_L] \bigr] &\eIMPEN a\cr
    &  \;\;\;\;\;\;\;\;\;
    - S_{\rm pert}^{(1,1)} \bigl[ \cL_L + F_M^{(0,1)}[\cL_L] +
            \delta \cL^{(0,1)} \bigr] &\eIMPEN b\cr
    &  \;\;\;\;\;\;\;\;\; - \int F_M^{(1,1)}[ \cL_L +
        \delta \cL^{(1,0)}] &\eIMPEN c\cr
  }
$$
This seems to express the matching information at this order in its
most impenetrable form, so we will examine the right hand side term
by term.
Though we will discuss examples
of only the instanton diagrams, the anti-instanton
diagrams are present as well.
Term \eIMPEN{a}\ is the
sum over all diagrams with either a heavy or light particle loop, and
with an instanton running over all scales (see \fEXPLONE).
Term \eIMPEN{b}\ is the sum over all diagrams with
a light particle loop and either with an instanton whose scale runs
from $0$ to $M$, or with the low energy approximation to an instanton
interaction where the instanton scale runs from $M$ to $\infty$ (see
\fEXPLTWO).  Finally, term \eIMPEN{c}\ is the sum
over diagrams with (the low energy approximation to) a heavy fermion
loop and an instanton whose scale ranges from $0$ to $M$ (see
\fEXPLTHREE).

\ifx\pansw\pictures
\medskip
$$
\beginpicture
\setcoordinatesystem units <\tdim,\tdim>
\setplotarea x from -60 to 40 , y from -55 to 35
\stpltsmbl
\put {\figfont $\underbrace{S_{\rm pert}^{(1,1)}
        \bigl[ \cL_H + \cL_L +
        \overbrace{ F_\infty^{(0,1)} [\cL_H + \cL_L] }^{
            \beginpicture
            \setcoordinatesystem units <\tdim,\tdim>
            \setplotarea x from -20 to 20 , y from 3 to 20
            \stpltsmbl
            \plot 0 0   0  65 /
            \plot 0 65  60 65 /
            \tarrow from 60 65 to 65 65
            \endpicture
        }
        \;\; \bigr]} _{
            \beginpicture
            \setcoordinatesystem units <\tdim,\tdim>
            \setplotarea x from -20 to 20 , y from -7 to -20
            \stpltsmbl
            \plot 0 0   0 -55 /
            \tarrow from 0 -55 to 0 -60
            \endpicture
        }$} at 0 0
    \put {\beginpicture
        \multiput {\jagdr}  at 0 20 *3 0 -10 /
        \multiput {\srd}  at -20 -20 *3 10 0 /
        \plot -20 20 20 20 /
        \put {$\bullet$} at 0 0
        \put {\figfont $< \infty$} at 15 0
        \put {\figfont $+\;\;\cdots$} [l] at 45 0
    \endpicture} [l] at 120 70
\put {\beginpicture
    \setcoordinatesystem units <\tdim,\tdim>
    \setplotarea x from -20 to 20 , y from -20 to 20
    \stpltsmbl
    \put {\beginpicture
         \put {\hfloop} at 0 0
         \multiput {\sru}  at -45 0 *2 10 0 /
         \multiput {\jagru} at 15 0 *3 10 0 /
         \put {$\bullet$} at 35 0
         \put {\figfont $< \infty$} at 35 10
         \plot 55 20 55 -20 /
         \endpicture} at -40 0
    \put {$+$} at 30 0
    \put {\beginpicture
         \put {\lfloop} at 0 0
         \multiput {\sru}  at -45 0 *2 10 0 /
         \multiput {\jagru} at 15 0 *3 10 0 /
         \put {$\bullet$} at 35 0
         \put {\figfont $< \infty$} at 35 10
         \plot 55 20 55 -20 /
         \endpicture} at 90 0
    \put {\figfont $+\;\;\cdots$} [l] at 155 0
\endpicture} [t] at 80 -80
\put { \baselineskip = 10pt \figfont
    Figure \xfig\fEXPLONE: Example diagrams for
    term \eIMPEN{a}.} at 55 -140
\endpicture
$$
\else\fi

\ifx\pansw\pictures
\medskip
$$
\beginpicture
\setcoordinatesystem units <\tdim,\tdim>
\setplotarea x from -60 to 40 , y from -55 to 35
\stpltsmbl
\put {\figfont
    $
        \underbrace{
            S_{\rm pert}^{(1,1)} \bigl[ \cL_L +
            \overbrace{
                F_M^{(0,1)}[\cL_L]
            }
            ^{
                \beginpicture
                \setcoordinatesystem units <\tdim,\tdim>
                \setplotarea x from -20 to 20 , y from 3 to 20
                \stpltsmbl
                \plot 0 0   0  65 /
                \plot 0 65  -15 65 /
                \tarrow from -15 65 to -20 65
                \endpicture
            }
            +
            \overbrace{
                \delta \cL^{(0,1)}
            }
            ^{
                \beginpicture
                \setcoordinatesystem units <\tdim,\tdim>
                \setplotarea x from -20 to 20 , y from 3 to 20
                \stpltsmbl
                \plot 0 0   0  65 /
                \plot 0 65  15 65 /
                \tarrow from 15 65 to 20 65
                \endpicture
            }
            \bigr]
        }
        _{
            \beginpicture
            \setcoordinatesystem units <\tdim,\tdim>
            \setplotarea x from -20 to 20 , y from -7 to -20
            \stpltsmbl
            \plot 0 0   0 -55 /
            \tarrow from 0 -55 to 0 -60
            \endpicture
        }
    $
} at 0 0
\put {\beginpicture
    \setcoordinatesystem units <\tdim,\tdim>
    \setplotarea x from -20 to 20 , y from -20 to 20
    \stpltsmbl
    \multiput {\jagdr}  at 0 20 *3 0 -10 /
    \multiput {\srd}  at -20 -20 *5 10 0 /
    \plot -20 20 40 20 /
    \put {$\bullet$} at 0 0
    \put {\figfont $> M$} at 15 0
    \put {\figfont $+\;\;\cdots$} [l] at 55 0
    \multiput {.} at -10 -35 *16 0 4 /
    \multiput {.} at -10 33 *10 4 0 /
    \multiput {.} at 30 -35 *16 0 4 /
    \multiput {.} at -10 -35 *10 4 0 /
    \endpicture
} [l] at 75 70
\put {\beginpicture
    \setcoordinatesystem units <\tdim,\tdim>
    \setplotarea x from -20 to 20 , y from -20 to 20
    \stpltsmbl
    \multiput {\jagdr}  at 0 20 *3 0 -10 /
    \multiput {\srd}  at -20 -20 *3 10 0 /
    \plot -20 20 20 20 /
    \put {$\bullet$} at 0 0
    \put {\figfont $< M$} at 15 0
    \put {\figfont $+\;\;\cdots$} [l] at 35 0
    \endpicture
} [r] at -45 70
\put {
    \beginpicture
    \setcoordinatesystem units <\tdim,\tdim>
    \setplotarea x from -20 to 20 , y from -20 to 20
    \stpltsmbl
    \put{ \beginpicture
        \put {\lfloop} at 0 0
        \multiput {\sru}  at -45 0 *2 10 0 /
        \multiput {\jagru} at 15 0 *3 10 0 /
        \put {$\bullet$} at 35 0
        \put {\figfont $< M$} at 35 10
        \plot 55 20 55 -20 /
        \endpicture
    } at -40 0
    \put {$+$} at 30 0
    \put {\beginpicture
        \put {\lfloop} at 0 0
        \multiput {\sru}  at -45 0 *2 10 0 /
        \multiput {\jagru} at 15 0 *3 10 0 /
        \put {$\bullet$} at 35 0
        \put {\figfont $> M$} at 35 10
        \plot 55 25 55 -25 /
        \multiput {.} at 6  16 *14 4 0 /
        \multiput {.} at 6 -16 *14 4 0 /
        \multiput {.} at  6 16 *7 0 -4 /
        \multiput {.} at  62 16 *7 0 -4 /
        \endpicture
    } at 90 0
    \endpicture
} [t] at 80 -80
\put { \baselineskip = 10pt \figfont
    Figure \xfig\fEXPLTWO: Example diagrams for
    term \eIMPEN{b}.} at 55 -140
\endpicture
$$
\else\fi

\ifx\pansw\pictures
\medskip
$$
\beginpicture
\setcoordinatesystem units <\tdim,\tdim>
\setplotarea x from -60 to 40 , y from -55 to 35
\stpltsmbl
\put {\figfont
    $
        \underbrace{
            \int F_M^{(1,1)}[ \cL_L +
            \overbrace{
                \delta \cL^{(1,0)}
            }
            ^{
                \beginpicture
                \setcoordinatesystem units <\tdim,\tdim>
                \setplotarea x from -20 to 20 , y from 3 to 20
                \stpltsmbl
                \plot 0 0   0  65 /
                \plot 0 65  25 65 /
                \tarrow from 25 65 to 30 65
                \endpicture
            }
            ]
        }
        _{
            \beginpicture
            \setcoordinatesystem units <\tdim,\tdim>
            \setplotarea x from -20 to 20 , y from -7 to -20
            \stpltsmbl
            \plot 0 0   0 -40 /
            \tarrow from 0 -40 to 0 -45
            \endpicture
        }
    $
} at 0 0
\put {\beginpicture
    \put {\lebox} at -1 0
    \put {\hfloop} at 0 0
    \multiput {\sru}  at -45 0 *2 10 0 /
    \multiput {\sru}  at 15 0 *2 10 0 /
    \put {\figfont $+\;\;\cdots$} [l] at 55 0
    \endpicture
} [l] at 70 70
\put {\beginpicture
    \put {\beginpicture
        \put {\hfloop} at 0 0
        \put {\lebox} at -1 0
        \multiput {\sru}  at -45 0 *2 10 0 /
        \multiput {\jagru} at 15 0 *3 10 0 /
        \put {$\bullet$} at 35 0
        \put {\figfont $< M$} at 35 10
        \plot 55 20 55 -20 /
        \endpicture
    } at 0 0
    \put {$+\;\;\cdots$} at 75 0
    \endpicture
} [t] at 0 -80
\put { \baselineskip = 10pt \figfont
    Figure \xfig\fEXPLTHREE: Example diagrams for
    term \eIMPEN{c}.} at 55 -140
\endpicture
$$
\else\fi

\medskip
Then, for $\int \delta \cL^{(1,1)}$,  equation \eIMPEN{}\ gives the
diagrams indicated in \fDLONEONE.
\ifx\pansw\pictures
$$
\beginpicture
\setcoordinatesystem units <\tdim,\tdim>
\setplotarea x from -60 to 40 , y from -55 to 35
\stpltsmbl
\put {\figfont $\int \delta \cL^{(1,1)} =$} [r] at -105 0
\put {\beginpicture
         \put {\hfloop} at 0 0
         \multiput {\sru}  at -45 0 *2 10 0 /
         \multiput {\jagru} at 15 0 *3 10 0 /
         \put {$\bullet$} at 35 0
         \put {\figfont $< \infty$} at 35 10
         \plot 55 33 55 -33 /
         \endpicture} at -40 5
\put {$-$} at 30 0
\multiput {.} at -70   22 *58 4 0 /
\multiput {.} at -70  -22 *58 4 0 /
\multiput {.} at -70   22 *10 0 -4 /
\multiput {.} at  162  22 *10 0 -4 /
\put {\beginpicture
         \put {\hfloop} at 0 0
         \multiput {\sru}  at -45 0 *2 10 0 /
         \multiput {\jagru} at 15 0 *3 10 0 /
         \put {$\bullet$} at 35 0
         \put {\figfont $< M$} at 35 10
         \plot 55 33 55 -33 /
         \put{\lebox} at -1 0
         \endpicture} at 100 0
\put {$+$} [l] at -120 -60
\put {\beginpicture
         \put {\lfloop} at 0 0
         \multiput {\sru}  at -45 0 *2 10 0 /
         \multiput {\jagru} at 15 0 *3 10 0 /
         \put {$\bullet$} at 35 0
         \put {\figfont $> M$} at 35 10
         \plot 55 33 55 -33 /
         \endpicture} at -50 -55
\put {$-$} at 20 -60
\multiput {.} at -80   -38 *61 4 0 /
\multiput {.} at -80   -82 *61 4 0 /
\multiput {.} at -80   -38 *10 0 -4 /
\multiput {.} at  164  -38 *10 0 -4 /
\put {\beginpicture
         \put {\lfloop} at 0 0
         \multiput {\sru}  at -45 0 *2 10 0 /
         \multiput {\jagru} at 15 0 *3 10 0 /
         \put {$\bullet$} at 35 0
         \put {\figfont $> M$} at 35 10
         \plot 55 33 55 -33 /
         \multiput {.} at 6  16 *14 4 0 /
         \multiput {.} at 6 -16 *14 4 0 /
         \multiput {.} at  6 16 *7 0 -4 /
         \multiput {.} at  62 16 *7 0 -4 /
         \endpicture} at 100 -60
\put {$+\;\;\cdots$} [l] at -120 -95
\put { \baselineskip = 10pt \figfont \lines {
    Figure \xfig\fDLONEONE: Matching contribution at (1,1). \cr
    For each instanton diagram, there is \cr
    a corresponding anti-instanton diagram. \cr}} at 0 -120
\endpicture
$$
\else\fi

The grouping of terms in \fDLONEONE\ is meant to be suggestive, in
that low energy approximation boxes have been drawn
around pairs of diagrams whose
IR physics cancels within each pair.  It should be stressed that a low
energy box merely indicates the finite order momentum expansion
through which $\delta \cL^{(l,d)}$ is defined, so that the same result
for $\delta \cL^{(l,d)}$ is indicated whether a single box is drawn
around the complete set of diagrams contributing to
$\delta \cL^{(l,d)}$ or several are drawn, one around each element in
any allowed partition of the set of diagrams.  A partition is only
allowed, however, if each element is IR analytic, otherwise the
momentum expansion of that element fails.  Fig. \xfig\fDLONEONE\ is
meant to suggest a partition of the set of diagrams for which each
element consists of the fewest allowable number of diagrams.
In the first low energy box in \fDLONEONE, the
first term represents diagrams
consisting of a heavy fermion loop with an instanton running over all
scales.  From these are subtracted diagrams represented by
the second term:  each is nominally the same diagram as one
indicated in the
first term, except the heavy fermion loop has been replaced by its low
energy approximation,
and the instanton scales run only from 0 to $M$.  To
this must be added the diagrams represented those in the second low
energy box: the first term represents
all diagrams consisting of a
light fermion loop and an instanton whose scale runs from $M$ to
$\infty$.  From this must be subtracted the diagrams
represented by the second term: each is nominally the same diagram as
one in the first term, except that the instanton-background propagators
have been replaced by their low energy approximation.

We have seen that in all the terms of $\delta \cL$ up to $\cO(1,1)$,
the essential features of the second section have continued to hold.
All infrared divergences, including those coming from large
instantons, cancel out of the matching calculation.
This is most interesting in the expression for $\delta \cL^{(1,1)}$.
For example in the first low energy box in \fDLONEONE, the large
instantons cancel
because the larger the instanton, the better the approximation to the
heavy fermion loop in the second term.  It should also be stressed
that, for example, the first and second terms
of the second low energy box do not cancel exactly
when the momentum expansion is carried out only to some finite order.

\newsec{Practical Calculations}

For a given term induced by instantons in the low energy theory, there
are only a finite number of diagrams to calculate at each order in
$(l,d)$.  It is unlikely that all of these terms could be calculated
analytically, however, because some of them involve the propagation of
massive particles in an instanton background.

In nonabelian gauge theory, such a propagator for a massless particle of
arbitrary spin and isospin can be calculated in closed form
\ref\rBCCL{L.S. Brown, R.D. Carlitz, D.B. Creamer, and C. Lee, Phys.
Lett. {\bf 70B} (1977) 180;
Phys. Lett. {\bf 71B} (1977) 103;
Phys. Rev. {\bf D17} (1978) 1583.}.
Also, the corresponding massive propagator
has been related to that of a scalar
\ref\rBROWNLEE{L.S. Brown and C. Lee, Phys. Rev. {\bf D18} (1978)
2180.}.  However, we believe the scalar
massive propagator has not been calculated in closed form.

In addition, the requirement that the quantum fluctuations about the
instanton in the zero-mode direction be treated specially forced us to
non-trivially modify the propagator of the field in which the
instanton is found (see equation \eMODPROP).  However, it is quite
possible that this new propagator could still be calculated in closed
form using methods similar to \rBCCL.

Whatever difficulties there are in exact analytic calculations, these
diagrams could perhaps be calculated numerically or estimated
analytically in certain regimes.

Finally, it seems likely that, in many calculations, instanton effects
in the low energy theory will dominate any of the instanton induced
matching corrections described here.  This is especially true in a
classically scale invariant (in the bosonic sector),  asymptotically
free theory like QCD, where infrared dominance seems to be a feature
of most instanton calculations.  We feel that it is nonetheless useful
to have a method for systematically calculating the corrections, both
for conceptual reasons and because they may sometimes be important.

\bigbreak\bigskip\bigskip\centerline{{\bf Acknowledgements}}\nobreak
STO wishes to thank Steve Hsu and Shane Hughes for useful discussions. This
work was supported in part by the National Science Foundation, under grant
\#PHY-9218167, and in part by the Texas National Research Laboratory
Commission under grant \#RGFY93-278B. The original idea for this work grew out
of discussions between HG and Guido Martinelli and others at the Institute for
Theoretical Physics in Santa Barbara.  HG is grateful to the staff of the ITP
for its hospitality and help.
\listrefs
\bye